\documentclass[pre,nofootinbib,twocolumn,floatfix]{revtex4}
\usepackage{graphicx}
\usepackage{amsmath}
\bibpunct[, ]{[}{]}{,}{a}{,}{,}
\ifnum\lefthyphenmin<2\lefthyphenmin=2\fi
\ifnum\righthyphenmin<2\righthyphenmin=2\fi

\begin{document}

\title{Analytical Solution of a Stochastic Content Based Network Model}

\author{Muhittin Mungan$^{1}$, Alkan Kabak\c c\i o\u glu$^2$, 
Duygu Balcan$^{3}$, Ayse Erzan$^{3,4}$}
\affiliation{$^1$Department of Physics, Faculty of Arts and Sciences \\
Bo\u gazi\c ci University, 34342 Bebek Istanbul, Turkey}
\affiliation{$^2$Dipartimento di Fisica, Universit\`a di Padova, I-35131 Padova, Italy}
\affiliation{$^3$Department of Physics, Faculty of Sciences and 
Letters\\
Istanbul Technical University, Maslak 34469, Istanbul, Turkey}
\affiliation{$^4$G\"ursey Institute, P.O.B. 6, \c Cengelk\"oy, 34680 Istanbul, Turkey}

\date{\today}
\begin{abstract}
We define and completely solve a content-based directed network whose nodes consist of 
random words and an adjacency rule involving perfect or approximate matches, for an alphabet 
with an arbitrary number of letters. The analytic expression for the out-degree 
distribution shows a crossover from a leading power law behavior to a 
log-periodic regime bounded by a different power law decay. The leading 
exponents in the two regions have a weak dependence on the mean word length, 
and an even weaker dependence on the alphabet size. The in-degree 
distribution, on the other hand, is much narrower and does not 
show scaling behavior. The results might be  of interest for understanding 
the emergence of genomic interaction networks, which rely, to a 
large extent, on mechanisms based on sequence matching, and exhibit 
similar global features to those found here.
 

\end{abstract}

\maketitle

\section{Introduction}

In a previous paper, two of us (Balcan and Erzan)~\cite{Balcan-Erzan} introduced  and 
numerically simulated a content based network~\cite{Vespignani}, 
with random binary strings associated with each node. The network arose 
by postulating a directed edge to exist between the nodes $i$ and $j$, 
if and only if the string, which can be regarded as a random word 
associated with the $i$th node, occurred at least once in the random 
word associated with the $j$th. 

This stochastic network was shown~\cite{Balcan-Erzan} to display 
distinctly different topology than either the classical random networks 
of Erd\"os and Renyi~\cite{Erdos} or the ``scale free" networks of the 
preferential-attachment universality class, introduced by 
Barabasi and Albert~\cite{Barabasi1999,Barabasi}.
Simulations~\cite{Balcan-Erzan} revealed that 
the in- and out-degree distributions, were markedly different, 
with in-degree distribution being rather localised.
The out-degree distribution displayed a sharp crossover behavior. 
For small out-degree $d$, the distribution $n(d)$ exhibited a putative 
scaling behavior over a very narrow region, where the log-log plot could 
be fitted with a straight line with slope $-\gamma_1 \simeq -1$, whereas, 
for larger $d$, log-periodic oscillations were found, with an envelope 
which could again be fitted, on a double logarithmic plot, by a linear 
graph with a slope $- \gamma_2\simeq - 1/2$. 

The purpose of this paper is twofold. We first extend 
the model of Balcan and Erzan~\cite{Balcan-Erzan} to 
a broader class of models in which the random strings are derived from an $r+1$ letter 
alphabet and where partial matches are allowed. Second, we obtain analytical 
expressions for the  ensemble averaged in- and out-degree distributions 
and investigate the crossover behavior of the out-degree distribution.  
We show that the putative scaling behavior observed 
in the simulations to coincides with the leading power law behavior 
obtained from our analytical results.
We describe in detail the finite size corrections to the infinite network limit.
Comparison of our analytical predictions with the numerical data~\cite{Balcan-Erzan}  for 
the $r=2$ random bit string model with perfect matches shows very good agreement. 

The paper is organized as follows: In the next section we reformulate the random string 
model of~\cite{Balcan-Erzan} for an alphabet of $r+1$ letters. Our analytical results depend 
on  the matching probability $p(l,k)$ that a string of length $l$ selected randomly from the 
set of all strings of length $l$ is contained at least once in a string of length $k$, $k 
\ge l$, that has been selected randomly from the set of all strings of length $k$. In 
Section III we derive an approximate form for this probability that is valid for moderately 
long strings $k \lesssim r^l$ and that allows for partial matches. Using the results of 
Section III, we obtain in Section IV analytical expressions for the in- and 
out-degree 
distributions. We investigate the scaling behavior of the out-degree distribution in these 
models and compare our results with the numerical data 
of~\cite{Balcan-Erzan}. We conclude this paper with a 
discussion of the possible relevance of 
our results to genomic networks, in Section V.
     
\section{The Random String Model}

Consider a random sequence  $C$ of fixed length $L$, consisting 
of letters from an alphabet $A$ 
of $r+1$ letters. The elements of the sequence $C$, 
$x \in \{0,1, \ldots, r\}$ are assumed to be independently 
and identically distributed according to
\begin{equation}
P(x) = p \delta(x-r) + (1-p)\frac{1}{r}\sum_{m=0}^{r-1}\delta(x-m)\;\;.
\end{equation}

A subsequence $G_i$ of $C$,  composed of 
the letters  $\{0, \ldots, r-1\}$ only, sandwiched between 
the $i^{th}$ and $(i+1)^{th}$ occurrences of the letter ``$r$,'' 
will be denoted the $i$th ``random word," or ``string," and will be 
associated with the $i$th vertex of a graph. 
For convenience, we assume that a letter ``$r$'' has also been placed 
at the $0^{th}$ and $(L+1)^{th}$ positions. 
With these definitions, the $i^{th}$ string can be written,
\begin{equation}
G_i = x_{i,1},x_{i,2},\ldots,x_{i,\ell_i}, \ \ i = 1,2,\ldots,N 
\end{equation}  
where $N$ is the number of strings (equivalently,  vertices), 
the ``letter'' $x_{i,\lambda} \in \{0,r-1\}$, 
$\lambda = 1,\ldots,\ell_i$, and $\ell_i$ is the length of the $i^{th}$ 
string $G_i$. Let $n_\ell$ be the number of strings of length $\ell$ and 
$q = 1 -p$. It follows that
\begin{equation}
\label{sum_rules}
\sum_i{\ell_i} = L-N\ , \ \ \ \sum_{\ell}{n_\ell} = N\ ,
\end{equation}
\begin{equation}
\label{averages}
\langle \ell \rangle = p^{-1} -1, 
 \ \ \ \langle n_\ell \rangle = Lp^2q^\ell, \  \ \ 
\langle N \rangle = Lp .
\end{equation}
Unless noted otherwise, we will assume that 
$L$ and $Lp$ are sufficiently large so that fluctuations in the number and 
length of the strings for different realizations of the
random sequence $C$ can be neglected when calculating statistical 
properties of quantities of interest. We will also discard the cases with $\ell = 0$ 
and construct the  graph from the remaining
vertices. The adjacency matrix is defined by the matching
condition
\begin{equation}
\label{eqn:wij}
w_{ij} = 
\begin{cases}
  1& G_i \subset G_j,\\
  0& \text{otherwise}.
\end{cases}
\end{equation}
By $G_i \subset G_j$ we mean that there exists an integer $\lambda$ such that
$0 \le \lambda \le \ell_j - \ell_i$ and
\begin{equation}
x_{i,l} = x_{j,\lambda + l}, \ \ l = 1,\ldots,\ell_i. 
\label{eqn:perfectmatch} 
\end{equation}
Two vertices are said to be connected
if the string $G_i$ appears as a subsequence of $G_j$, or
 in other words $G_j$ {\em matches} $G_i$. Thus $w_{ij}=1$ 
indicates a directed link (an edge) from $G_i$ to $G_j$. We
will also consider {\em imperfect} matches, where Eq.~(\ref{eqn:perfectmatch}) is
valid only for some values of $l$ rather than all values. 
In order to avoid ambiguity we will refer to the former case as a {\em perfect} match.
For $Lp$ large enough ($p>p_c(L)$, see~\cite{Balcan-Erzan}), which is assumed here, the 
graph consists of one giant cluster. We will henceforth refer to this graph 
as the network, and denote the vertices, or equivalently, 
the strings associated with them, as the ``nodes."

\begin{figure}[h!]
\begin{center}
\end{center}
\includegraphics[width=8cm]{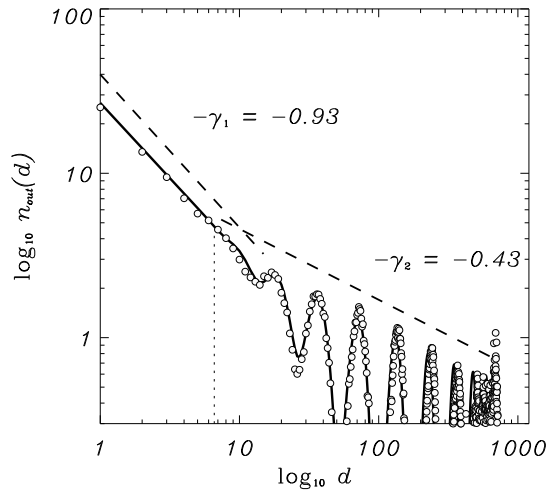}
\caption[]{Scaling behavior of the out-degree distribution. 
The numerical data (circles) shows a cross-over in the scaling behavior
from small values of the out-degree to larger values. The solid line
is the theoretical expression. The dashed
lines serve as a guide to the eye for the predicted scaling behavior
and have been offset for clarity. The cross-over occurs at $d_c= 6.6$ and has
been shown as a vertical line.  }
\label{outdistscaling}
\end{figure}

The resulting network was numerically studied earlier by Balcan and Erzan in
\cite{Balcan-Erzan}, for the case of binary strings, i.e., $r = 2$, 
and perfect matches Eq.~(\ref{eqn:perfectmatch}), where it was shown that 
the logarithm of the out-degree distribution behaved linearly 
over a very narrow, initial range, 
with a slope of $\simeq -1$.  Beyond a crossover point the distribution 
exhibited an oscillatory behavior, whose envelope again behaved 
linearly on a log-log plot, with a different slope, namely $\simeq -1/2$.
The  out-degree distribution is shown in Figure (\ref{outdistscaling}), where 
the numerical results were obtained~\cite{Balcan-Erzan} 
 by averaging the out-degree distributions 
over 500 graphs, associated with independently generated sequences of length $L=15000$, 
and $p = 0.05$.
Notice the strong oscillatory behavior. It turns out that each peak in 
the out-degree distribution is supported predominantly by the  
out-degrees of genes with a corresponding common length $l$. 

\vspace*{1cm}

In order to proceed with the analytical treatment, it is convenient to
group the $G_i$ into subsets according to their lengths and we define
\begin{equation} 
{\cal G}_l = \{ G_i | \ell_i = l\}\ .
\end{equation}
It turns out that that the central quantity determining the behavior of the 
in- and out-degree distributions is the probability $p(l,k)$
that a string in ${\cal G}_l$  has an outgoing edge terminating in a 
member of ${\cal G}_k$. We therefore turn next to the derivation of $p(l,k)$. 
The discussion of the degree distributions will then be 
taken up in Section IV.  
 
\section{Analytical Results for the Matching Probability}

Let $x$, and $y$ be variables such that
$x, y, \in \{0, \ldots , r-1\} $. 
Define an interaction $u(x,y)$ between $x$ and $y$ as

\begin{equation} 
u(x,y) = 1 - \delta (x-y). 
\label{eqn:udef}
\end{equation} 
Let ${\bf x} = (x_1,x_2,x_3, \ldots, x_{l})$ and 
${\bf y} = (y_1,y_2,y_3, \ldots, y_{l})$,
be two strings of $l$ letters and define their interaction 
$ U({\bf x},{\bf y})$ as
\begin{equation}
U({\bf x},{\bf y }) = \sum_{t=1}^{l} u(x_t,y_t).
\end{equation}
The function $U({\bf x}, {\bf y})$, as defined above, counts the
number of unmatched letters between strings ${\bf x}$ and $ {\bf y}$. 

Introduce an ``inverse temperature'' $\beta$ and consider the 
Boltzmann factor $e^{-\beta U}$. In the ``zero temperature'' limit we
have
\begin{equation}
\lim_{\beta \rightarrow \infty} e^{-\beta U({\bf x}, {\bf y})} = 
\left \{\begin{array}{ll}  1, & \mbox{if ${\bf x} = {\bf y}$} \\
                                      0,   & \mbox{otherwise}. \end{array} \right. 
\end{equation}
We see that the limit $\beta \rightarrow \infty $ is a ``no tolerance'' limit
\cite{Ozcelik}, enforcing perfect matching of {\bf x} and {\bf y}, {\em i.e.} 
$x_t = y_t$, $t = 1, 2, \ldots, l$.
Let ${\bf y} = (y_1,y_2, \ldots , y_k) $ be a string of length 
$k \ge l$ and denote by ${\bf y}_{a,l} = ( y_{a+1}, y_{a+2}, \ldots , y_{a+l})$
the substring of length $l$ starting at position $a$, $ a = 0, 1, \ldots, k-l$. 
Furthermore let
\begin{equation}
f_a({\bf x},{\bf y};\beta) = e^{-\beta U({\bf x},{\bf y}_{a,l})}.   
\label{eqn:fdef}
\end{equation}
so that we have
\begin{equation}
f_a({\bf x},{\bf y}) \equiv \lim_{\beta \rightarrow \infty}  f_a({\bf x},{\bf y};\beta)
 = \left \{\begin{array}{ll}  1, & \mbox{$ {\bf x} = {\bf y}_{a,l}$ } 
          \\ 0,   & \mbox{otherwise}. \end{array} \right.
\label{eqn:fadef}
\end{equation}
Thus, $f_a({\bf x},{\bf y}) = 1$,
if and only if ${\bf x}$ matches ${\bf y}$ at position $a$, and zero otherwise.

Likewise, let $f({\bf x},{\bf y})$ be a function that takes on the value
one if the $k$-string ${ \bf y}$ contains the given $l$-string ${\bf x}$ and
zero otherwise. Note that the complement of 
the event that  ${\bf x}$ matches ${\bf y}$ is the event that ${\bf x}$ 
does not match ${\bf y}$ anywhere. Thus, using Eq.~(\ref{eqn:fadef}), we can write
\begin{equation}
f({\bf x},{\bf y}) = 1 -  \prod_{a=0}^{k-l}
\left [ 1 - f_a({\bf x},{\bf y}) \right ].
\end{equation}
Letting $p(l,k; { \bf x})$ denote the probability  that a randomly drawn $k$-string ${ \bf y}$ contains a given $l$-string ${\bf x}$, we therefore find
\begin{equation}
p(l,k; { \bf x}) = 1 - \frac{1}{r^k} \sum_{{\bf y}}^{} \prod_{a=0}^{k-l}
\left [ 1 - f_a({\bf x},{\bf y}) \right ],
\label{eqn:p1function}
\end{equation}
where $r^k$ is the number of distinct $k$-strings of $r$-letters, 
and $\sum_{{\bf y}}^{}$ denotes the sum over all such strings ${\bf y}$.

Generalizing the above equation to incorporate partial matches we obtain: 
\begin{equation}
p(l,k;{\bf x}) = \lim_{\beta \rightarrow \infty} p(l,k;{\bf x},\beta),
\end{equation}
where
\begin{equation}
p(l,k;{\bf x},\beta) = 1 - \frac{1}{r^k} \sum_{{\bf y}}^{} \prod_{a=0}^{k-l}
\left [ 1 - f_a({\bf x},{\bf y};\beta) \right ].
\label{eqn:pfunction}
\end{equation}

The products in equation (\ref{eqn:pfunction}) can be expanded and we 
obtain a Mayer-like sum
\begin{eqnarray}
p(l,k;{\bf x},\beta) &=& \frac{1}{r^k} \sum_{{\bf y}}^{} \sum_{a} f_a  
 - \frac{1}{r^k} \sum_{{\bf y}}^{} \sum_{a < b} f_a f_b \nonumber \\
  &+& \frac{1}{r^k} \sum_{{\bf y}}^{} \sum_{a < b < c} f_a f_b f_c - \ldots,
\label{eqn:meier}
\end{eqnarray}
which we can write as
\begin{eqnarray} 
p(l,k;{\bf x},\beta) &=&  \sum_{a} W^{(1)}(a;{\bf x})  
 -  \sum_{a < b} W^{(2)}(a,b;{\bf x}) \nonumber \\
  &+&  \sum_{a < b < c} W^{(3)}(a,b,c;{\bf x}) - \ldots,
\label{eqn:meierdef}
\end{eqnarray}
where
\begin{eqnarray}
W^{(1)}(a;{ \bf x}) &=& \frac{1}{r^k} 
\sum_{{\bf y}}^{}f_a({\bf x},{\bf y};\beta) \nonumber \\
W^{(2)}(a,b;{ \bf x}) &=& \frac{1}{r^k} 
\sum_{{\bf y}}^{}f_a({\bf x},{\bf y};\beta)f_b({\bf x},{\bf y};\beta) \nonumber\\
W^{(3)}(a,b,c;{ \bf x}) &=& \frac{1}{r^k} 
\sum_{{\bf y}}^{}f_a({\bf x},{\bf y};\beta)f_b({\bf x},{\bf y};\beta)
f_c({\bf x},{\bf y};\beta) 
\nonumber \\ &\cdots& 
\label{eqn:meierterms}
\end{eqnarray}

Using equations (\ref{eqn:udef}) and (\ref{eqn:fdef}), we obtain 
\begin{equation}
W^{(1)}(a;{ \bf x}) = 
\frac{1}{r^l}\left [ 1 + (r-1) e^{-\beta } 
\right ]^{l} \equiv W^{(1)}.
\label{eqn:fi}
\end{equation}
Note that $W^{(1)}(a;{\bf x})$ is independent of $a$, and ${\bf x}$. 

Let us now turn to the second order term, $W^{(2)}(a,b;{\bf x})$ 
in Eqs.~(\ref{eqn:meierdef}) 
and (\ref{eqn:meierterms}). Here, we need to distinguish two cases, 
(i) $b - a \ge l$ and (ii) $b-a < l$.

In case (i), the set of indices of ${\bf y}_{a,l}$ and ${\bf y}_{b,l}$ are 
distinct and the evaluation of the partition sum proceeds 
analogously to equation (\ref{eqn:fi}) yielding
\begin{equation}
W^{(2)}(a,b;{\bf x}) = 
\left (\frac{1}{r^l} \right )^2 \left [ 1 + (r-1)e^{-\beta} \right ]^{2l}, \, 
|b-a| \ge l.  
\label{eqn:fij1}   
\end{equation}

In case (ii), $|b-a| < l$, there is an overlap between the indices of 
${\bf y}_{a,l}$ and ${\bf y}_{b,l}$. Letting $|b-a| = m$, we find
\begin{eqnarray}
W^{(2)}(a,b;{\bf x}) =   
\frac{1}{r^{l+m}} \left [ 1 + (r-1)e^{-\beta} \right ]^{2m} \nonumber \\
\times \prod_{t=1}^{l-m}
   \left [ 1 + (r-1) e^{-2\beta} - u(x_t,x_{m+t})\left ( 1 - e^{-\beta } \right )^2 
   \right ], \nonumber \\
|b-a| < l.  
\label{eqn:fij2}
\end{eqnarray}
Note that $W^{(2)}(a,b;{\bf x})$, as defined Eqs.~(\ref{eqn:fij1}) and 
(\ref{eqn:fij2}), depends on ${\bf x}$ only when $|b-a| < l$. 
Next, we perform
the ${\bf x}$ average of  $W^{(2)}(l,k;\bf{x})$,
\begin{equation}
\frac{1}{r^l} \sum_{{\bf x}} W^{(2)}(a,b;{\bf x}) = 
  \frac{1}{r^{2l}} \left [ 1 + (r-1) e^{-\beta} \right ]^{2l}.
\label{eqn:factor1}
\end{equation}
The calculations leading to Eqs. (\ref{eqn:fij2}) and (\ref{eqn:factor1})
are a little involved and can be found in the appendix. 

Comparing 
Eqs.~(\ref{eqn:fi}) and (\ref{eqn:factor1}), we see that once averaged 
over ${\bf x}$, $W^{(2)}$ factorizes as
\begin{equation}
W^{(2)} = \left < W^{(2)}(a,b;{\bf x}) \right >_x = \left ( W^{(1)} \right )^2,
\label{eqn:factor2}
\end{equation}
or equivalently,
\begin{equation}
\left < f_{a} f_{b} \right >_{y,x} = 
 \left < f_{a} \right >_{y,x} \left < f_{b} \right >_{y,x}, \ \ \ a \ne b,
\end{equation}
where, for simplicity, we have introduced the short hand notation
$\left < \ldots \right >_{y,x} $ to denote averaging first over 
${\bf y}$ then ${\bf x}$.

Let us therefore make the approximation that all higher moments factorize similarly,
\begin{equation}
 \left < f_{a_1} f_{a_2} \ldots f_{a_s} \right >_{y,x} \simeq 
 \left < f_{a_1} \right >_{y,x} \left < f_{a_2} \right >_{y,x}
  \ldots \left < f_{a_s} \right >_{y,x}, 
\label{eqn:factorn}
\end{equation}
with $\{a_s\}$ being distinct. It can be readily shown that Eq.~(\ref{eqn:factorn})
is exact when $a_{i+1} - a_{i} > l$, i.e, there are no overlaps between the
segments at position $a_i$. 
Upon substituting Eq. (\ref{eqn:factorn})
into Eq. (\ref{eqn:meier}) and performing the ${\bf x}$ average we obtain the
matching probability
\begin{equation}
p(l,k;\beta) = \left < p(l,k;{\bf x},\beta) \right >_x,
\end{equation}
with
\begin{equation}
p(l,k;\beta) = 1 - 
 \left ( 1 -  \frac{1}{r^l} \left [ 1 + (r-1) e^{-\beta} \right ]^{l} 
           \right )^{k-l+1}.
\label{eqn:pijbeta}
\end{equation}
In the ``zero temperature'' limit ($\beta \rightarrow \infty$),
this becomes
\begin{equation}
p(l,k) = 1 - 
 \left ( 1 -  \frac{1}{r^l} 
           \right )^{k-l+1}.
\label{eqn:pijdef2}
\end{equation}
For $r^l \gg k$, $p(l,k;\beta)$ has  the asymptotic form
\begin{equation}
p(l,k;\beta) = 1 - 
 \exp { \left ( - \frac{k-l+1}{r^l} \left [ 1 + (r-1)e^{-\beta} \right ]^l \right )},
\end{equation}
which for $\beta \rightarrow \infty $ becomes
\begin{equation}
p(l,k) = 1 - 
 \exp {\left ( - \frac{k-l+1}{r^l} \right ) }.
\end{equation}
For very large $l$ this further reduces to
\begin{equation}
p(l,k) = {\frac{k-l+1}{r^l}}.
\label{eqn:naivepij}
\end{equation}

Note that a finite $\beta$ acts like an enhanced matching probability, i.e., a
false positive match. In the limit  $\beta \rightarrow 0$, the matching
probability becomes
\begin{equation} 
\lim_{\beta \rightarrow 0} p(l,k;\beta) = 1
\label{eqn:pijtinfty}
\end{equation} 
Hence the  ``high-temperature'' limit of our model 
corresponds to indiscriminate matches.

\vspace*{1cm}
\begin{figure}[h!]
\begin{center}
\end{center}
\includegraphics[width=8cm]{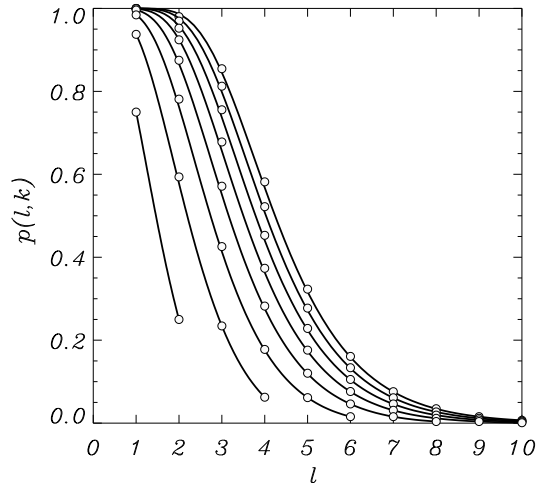}
\caption[]{Comparison of the exact matching probability
$p(k,l)$ (circles) with the approximate expression (\ref{eqn:pijdef2}) (lines) 
for $r=2$ and perfect matches. The 
curves are (from top to bottom) for values of $k = 16, 14, 12, 10, 8, 6, 4, 2$. }
\label{r2pdist}
\end{figure}
 
Of course, the crucial approximation, Eq.~(\ref{eqn:factorn}), 
is not correct in general and
one expects corrections coming from higher order correlations contained 
in Eq.~(\ref{eqn:meierdef}). 
These correlations are due to the fact that if a given 
string ${\bf x}$ is matched at a position $a$, this affects the likelihood 
of matching the same string at any nearby location $b$ with $\vert b-a \vert \lesssim l$. 
Nevertheless, the approximate result for $p(l,k)$, 
Eq.~(\ref{eqn:pijdef2}), 
is surprisingly good.  Fig. (\ref{r2pdist}) shows a comparison of the 
matching probability obtained from exact enumeration carried out computationally, 
with the analytical expression (\ref{eqn:pijdef2}) for $r=2$ and perfect
matches. As can be seen from the figure, there are only very small 
discrepancies for small $l$ when 
$k > 2^l$, e.g. data points around $k= 16, 14, 12$ with $l=4, 3, 2$.
Since our expression for $p(l,k)$, Eq.~(\ref{eqn:pijdef2}), is exact for $l=1$, 
there are no discrepancies at $l=1$. 

Notice that Eq.~(\ref{eqn:naivepij}) is the matching probability
that can alternatively be obtained by assuming the probabilities of 
matching a string of length $l$ at any position in a string of 
length $k$ are
independent, and equal, $1/r^l$. Eq.~(\ref{eqn:pijdef2}), 
on the other hand, is the matching probability that can also be found
assuming the probabilities of 
{\em not} matching a string of length $l$ at any position in a string 
of length $k$ 
are independent and equal, $1 - 1/r^l$. Thus the 
factorization approximation, Eq.~(\ref{eqn:factorn}),
leading to Eq.~(\ref{eqn:pijbeta}) implies that the probabilities of {\em not}
matching at a given position are independent. 

For the regime of interest, $k \lesssim 2^l$, this approximation leading
to Eq.~(\ref{eqn:pijdef2}) is extremely good. 
We think that this is due to the fact that the 
factorization property  underlying our approximation, Eq.~(\ref{eqn:factorn}), 
is exact for the two-point correlation function ($s=2$), Eq.~(\ref{eqn:factor2}).
This means that any corrections to this result must be coming from higher 
order correlations with strongly overlapping segments, since 
non-overlapping segments will factorize and thus reduce to lower order correlators.
This is very similar to the connected cluster expansion in 
statistical mechanics \cite{StatMechBooks}. 
Indeed, such an expansion can be set up, however the
calculations are rather tedious due to the discreteness of the 
problem and beyond the scope of this paper.   
Yet it is clear that the weight of an $s$-point correlation 
function with $s$ overlapping (connected) segments must be very small for
large $s$, since the overlap imposes very strong conditions on the 
structure of the string ${\bf x}$ to be matched.          

For the remainder of the paper it is convenient to define the quantities $t$
and $z$ as
\begin{eqnarray}
t &=& 1 -  \frac{1}{r^l} \left [ 1 + (r-1) e^{-\beta} \right ]^l = 
1 - z^l \label{eqn:tdef} \\
z &=&\frac{1}{r}\left [ 1 + (r-1) e^{-\beta} \right ], \label{eqn:zdef} 
\end{eqnarray}
where we have suppressed the $\beta$, $r$ and $l$ dependence 
for clarity.
Notice that the effect of the number of letters in the alphabet $r$ and
the extent of mismatch as parametrized by the ``inverse temperature'' 
$\beta$ enter into the expression for $p(l,k;\beta)$ as a single 
parameter, $z$, as defined above.
With the above definitions, Eq.~(\ref{eqn:pijbeta}) becomes
\begin{equation}
p(l,k;z) = 1 - t^{k-l+1} = 1 - \left ( 1 - z^l \right )^{k-l+1}.
\label{eqn:pdef2}
\end{equation}
The ``zero-temperature'' limit is given by $z = r^{-1}$, while the 
``high-temperature'' limit is $z = 1$. The range of $z$ is therefore, 
$z \in  (r^{-1},1)$, which for  $r \gg 1 $, approaches $z \in  (0,1)$.

We note in passing that the matching probability computed in this section, 
is in a sense complementary to the problem of sequence 
alignment~\cite{Karlin0,Karlin1}, which has important applications in the 
study of proteins and DNA.  The problem there is to identify subsequences of 
arbitrary length, showing strong similarity beyond pure statistical chance, 
{\it within} two long sequences sampling the same alphabet, possibly with 
different native probabilities. The pioneering work of Altschul, Karlin, 
et al.~\cite{Karlin0,Karlin1} yields a probability distribution for the 
similarity {\it score} of such likely regions, under the assumption that the 
region with the highest score is unique (i.e., non-degenerate), that the 
two sequences searched are of comparable length, and sufficiently long. 
The scoring scheme is to a large extent arbitrary as long as the 
scores corresponding to some degree of matching are rare (and positive) 
while those corresponding to mismatches are
much more probable (and negative). This arbitrariness may be removed 
by proper normalization 
and scores obtained via different schemes can be compared in a meaningful way.
The matching probability computed in the present paper could be related 
to the probability for the highest score (corresponding to an 
exact match without gaps), holding for the entire length of the 
shorter sequence.  However, our calculation makes no assumptions 
regarding the relative lengths of the two sequences, apart from the obvious 
requirement that $l \le k$.  The approximation to which we have to resort 
in the final solution works best when either the two sequences are 
almost of the same length, or if $k \lesssim  r^l$. Moreover there is no 
assumption regarding the number of times the highest score is achieved.  
More interestingly, the statistics of multiple high-scoring 
segments~\cite{Karlin2} could have been related to the out-degree 
statistics of a given node {\it had we taken each high scoring match 
in the complete random sequence to correspond to a different edge}. 
As it is, a single edge corresponds to the presence of one or more 
occurrences of a shorter string, say $G_i$, inside a longer string $G_j$. 
That is,  multiple occurrences of the shorter string within a subsequence 
of the complete random sequence are bunched together to result 
in a single edge between the nodes $i$ and $j$.  
 
We now turn to the calculation of the in- and out-degree distributions.

\section{The Degree Distributions}

In Section II we showed that the subsequences $\{G_{i}\}$ of a random sequence $C$
generate a network whose nodes are associated with these strings,  and whose edges are 
defined by the matching relation Eq.~(\ref{eqn:wij}). In this section
we will derive the in- and out-degree distribution associated with
this network. 

Consider a randomly selected string $G_i$. The in- and 
out-degree of the corresponding node, $d_{\rm in}(i)$ and 
$d_{\rm out}(i)$,  are defined by the total 
number of edges terminating in and originating 
from that node, respectively, 
\begin{eqnarray}
d_{\rm in}(i)  &=& \sum_{j} w_{ji} \nonumber \\ 
d_{\rm out}(i) &=& \sum_{j} w_{ij}.
\label{eqn:kdef}
\end{eqnarray}

The corresponding in- and out-degree distributions are given by
\begin{eqnarray}
n_{\rm in}(d)  &=& \sum_{i} \delta \left( d-d_{\rm in}(i) \right ) \nonumber \\
n_{\rm out}(d) &=& \sum_{i} \delta \left ( d-d_{\rm out}(i) \right ) .
\label{eqn:ndef}
\end{eqnarray}

\subsection{The Out-Degree Distribution}
Letting ${\cal G}_l$ denote the set of strings 
of length $l$, we can rewrite 
the out-degree distribution 
Eq.~(\ref{eqn:ndef}) as
\begin{equation}
n_{\rm out}(d) = \sum_{l=1}^{L}n_l \left [ \frac{1}{n_l} \sum_{j \in {\cal G}_l} 
\delta \left ( d-d_{\rm out}(j) \right ) \right]. 
\label{eqn:noutdef}
\end{equation}
For large $n_l$, the quantity in parentheses 
will approach the (conditional) probability $P_{\rm out}(X_l = d|l)$ that a randomly 
selected string whose length is given to be $l$ has an out-degree $d$. 

In the limit $L,N \rightarrow \infty$, such that $N/L = p$, the ratio of the 
number of strings, $N$, to the length of the whole random sequence, $L$, remains 
constant, all the possible $r^l$  realizations of random words of a given 
length $l$ will be present with equal respective weights and
we have, 
\begin{equation}
\lim_{L,N \rightarrow \infty} \frac{1}{n_l n_k}\sum_{i  \in {\cal G}_l}
\sum_{j \in {\cal G}_k} w_{ij} = p(l,k).
\label{eqn:plkdef}
\end{equation}
We will refer to this limit as the large-$L$ limit. 

The quantity $p(l,k)$, as defined in the above equation, is the probability that 
a randomly selected string of given length $l$ matches another independently and 
randomly selected string of length $k$. This probability has been calculated in 
Section III for the general case of imperfect matches, Eq.~(\ref{eqn:pijbeta}), as
well as perfect matches, Eq.~(\ref{eqn:pijdef2}). Eqs.~(\ref{eqn:noutdef}) and
(\ref{eqn:plkdef}) show the self-averaging property of the degree distribution in
the large-$L$ limit. 

Define the random variable $X_{l k}$, as the number of 
edges originating from a randomly selected string of length $l$ that 
terminate in strings of length $k$. Then $X_l$ can be written as a sum of 
the random variables $X_{l k}$, 
\begin{equation}
X_{l} = \sum_{k \ge l} X_{l k}.
\label{eqn:xioutsum}
\end{equation}

We can therefore write $X_l$ as 
\begin{equation}
\langle X_l \rangle = \frac{1}{n_l}\sum_{i \in {\cal G}_l}
\sum_{k=l}^{L}\sum_{j \in {\cal G}_k }w_{ij},
\end{equation}
or,   
\begin{equation}
\langle X_l \rangle  = \sum_{k=l}^{L} n_k 
\left[ \frac{1}{n_l n_k} \sum_{i \in {\cal G}_l} \sum_{j \in {\cal G}_k}  
 w_{ij} \right].
\label{eqn:xldef}
\end{equation}

We see from Eqs.~(\ref{eqn:xldef}) and 
(\ref{eqn:plkdef}) that in the large-$L$ limit
\begin{equation}
\left < X_{l k} \right > = n_k p(l,k),
\label{eqn:Xlkave}
\end{equation}
and
\begin{equation}
\left < X_l \right > = \sum_{k=l}^{L} n_k p(l,k),
\label{eqn:Xlave}
\end{equation}
where $\left < \cdots \right >$ denotes an average over all the strings of length $l$ 
in the complete random sequence. 
Note that in the large-$L$ limit $X_{l k}$ is binomially distributed,
\begin{equation}
P(X_{l k} = d | l) = \left ( \begin{array}{ll} n_k \\ d \end{array} \right ) p(l,k)^d 
\left ( 1 - p(l,k) \right )^{n_k - d }.
\label{eqn:pxij}
\end{equation}

As can be seen from Eq.~(\ref{eqn:xioutsum}), $X_l$ is
a sum of the random variables $X_{l k}$ and thus in the large-$L$ limit 
the central limit theorem assures that 
the distribution for $X_l$ will approach a Gaussian distribution, 
\begin{equation}
P_{\rm out}(X_l = d|l) = \frac{1}{\sqrt{2 \pi} \sigma_l } 
\exp \left[ -\frac{ ( d - d_l  )^2}{2 \sigma_l^2} \right],
\label{eqn:gauss}
\end{equation}
whose mean
$d_l$ and standard deviation $\sigma_l$ are given by those of $X_{l k}$, 
Eq.~(\ref{eqn:xioutsum}), according to: 
\begin{eqnarray}
d_l &=& \left < X_l \right > =  \sum_{k \ge l} \left < X_{l k} \right > 
\label{eqn:xiave}\\    
\sigma_{l}^{2} &=& \left < X_{l}^2 \right > - \left < X_l \right >^2 
= \sum_{k \ge l} \left < \sigma_{l k}^2 \right >,
\label{eqn:xisdev}
\end{eqnarray}
where
\begin{equation}
\sigma_{l k}^2 = \left < X_{l k}^2 \right > - \left < X_{l k} \right >^2 .
\end{equation}
For binomially distributed $X_{l k}$ we have
\begin{eqnarray}
\left < X_{l k} \right > &=& n_k p(l,k) \label{eqn:xelk} \\
\sigma_{l k}^2 &=& n_k p(l,k)\left ( 1 - p(l,k) \right ).
\end{eqnarray}
Using Eq. (\ref{eqn:pdef2}), one can readily carry out the sums in Eqs.
(\ref{eqn:xiave}) and (\ref{eqn:xisdev}) to find 
\begin{eqnarray}
d_l &=& \frac{N}{p+q z^l } \left ( qz \right )^l \label{eqn:di} \\
\sigma_{l}^2 &=& d_l \frac{pt}{1 - q t^2}. \label{eqn:sigsqi}
\end{eqnarray}
Noting also that the probability of selecting a string of 
length $l$ is $pq^l$, the total out-degree distribution
is given by
\begin{equation}
P_{\rm out}(d) = \sum_{l=1}^{L} pq^l P_{out}(X_l=d \vert l).
\label{eqn:poutgen}
\end{equation}
and thus in the large-$L$ limit we obtain
\begin{equation}
P_{\rm out}(d) = \sum_{l=1}^{L} pq^l 
\frac{1}{\sqrt{2 \pi} \sigma_l } 
\exp \left[ -\frac{ ( d - d_l  )^2}{2 \sigma_l^2} \right]
\label{eqn:poutgauss}
\end{equation}
with $d_l$ and $\sigma_l$ given by Eqs.~(\ref{eqn:di}) and (\ref{eqn:sigsqi}),
respectively.

As $l$ becomes large, $p(l,k)$ decreases towards zero. Thus
with increasing $l$ the binomial distribution of $X_{l k}$, Eq.~(\ref{eqn:pxij}), 
will approach a Poisson distribution of the same mean. Note that
the sum of independent and Poisson distributed 
random variables is also Poisson distributed with mean equal to the sum of the 
individual means. Thus for large $l$, $X_l$ as defined in Eq.~(\ref{eqn:xioutsum}),
is Poisson.  For a Poisson distributed random variable the variance equals to
its mean so that for large $l$ we expect
\begin{equation}
\sigma^2_l = d_l\;\;,
\label{also}
\end{equation}
as can also be directly verified by taking the appropriate 
limit in Eq.~(\ref{eqn:sigsqi}).

\subsubsection{Ensemble Averages and Finite Size Effects}

The numerical data of \cite{Balcan-Erzan} has been obtained from averaging over
500 realizations of a random sequence  of length $L=15000$ with $N=750$. A finite sample
size will cause sample to sample fluctuations in the number of strings, or ``random words." 
An average over a large ensemble of 
different realizations will yield the same average values for the out-degrees
as those obtained from a single random sequence of infinite length. However averaging
over many realizations will increase the fluctuations around the mean. It is not hard to 
see that this will affect predominantly nodes with large out-degrees, (short strings) 
where there is already self-averaging within the random sequence, 
but with a distribution which varies from sample to sample. 

Nodes with small out-degrees (long strings)
correspond to rare matches and thus for these nodes there is no self-averaging 
within the sample. To see this, consider the extreme case, where a sample 
contains on average one or less matches for such a node. When an ensemble 
average is taken, the dominant contribution to the 
variance of the out-degree will come from the sample to sample 
fluctuations.

Denoting the mean and variance of the out-degree of a node of length $l$, that 
has been corrected for the finite size, by $\tilde{d}_l$ and 
$\tilde{\sigma}^{2}_{l}$, respectively, we have
\begin{equation}  
\tilde{d}_l = d_l, \ \  \tilde{\sigma}^{2}_{l} \rightarrow \sigma^{2}_{l} 
\ \ \ \mbox{for large $l$}.
\label{eqn:largelsigma}
\end{equation}   
In what follows we will re-calculate previously introduced statistics, taking
into account the fluctuations in $n_k$. In order to avoid confusion, these quantities
will be denoted with a tilde.  

We can estimate $\tilde{\sigma}^{2}_{l}$ as follows. 
The random variable $\tilde{X}_{l k}$ itself is a sum of random variables:
\begin{equation}
\tilde{X}_{l k} = \sum_{j \in {\cal G}_k } Y_{lj},
\end{equation}
where  $Y_{lj} = 1$ if the string $G_j$ of length $k$ matches the (given) 
string of length $l$ 
and zero otherwise. Such an event constitutes a Bernoulli trial and its probability is
$p(l,k)$. The mean and variance of $Y_{lj}$ are given by
\begin{eqnarray}
\langle Y_{lj} \rangle &=&  p(l,k)
\label{eqn:yave} \\
\langle Y^2_{lj} \rangle - 
\langle Y_{lj} \rangle^2 &=&  p(l,k) (1- p(l,k))
\label{eqn:ysig}
\end{eqnarray}
The number of such trials is $\tilde{n}_k$, the number of elements of
${\cal G}_k$, and hence $\tilde{n}_k$ itself is a random variable. 
For sufficiently large $N$ and for
values of $\tilde{n}_k$ near the mean, the constraints, 
Eq.~(\ref{averages}), can be neglected and the probability 
of finding $\tilde{n}_k$ strings of length $k$ is approximately binomially distributed
\begin{equation}
P(\tilde{n}_k = n) =  \left ( \begin{array}{ll} N \\ n \end{array} \right ) 
\left (pq^k \right)^n \left ( 1 - pq^k \right )^{N - n }.  
\end{equation}
We thus find
\begin{eqnarray} 
  \langle \tilde{n}_k \rangle &=& Npq^k \label{eqn:nave}\\
 \tilde{\sigma}^2_{n_k} &=& pq^k (1 - pq^k) \label{eqn:nsig}.
\end{eqnarray}

\begin{figure}[h!]
\begin{center}
\end{center}
\includegraphics[width=8cm]{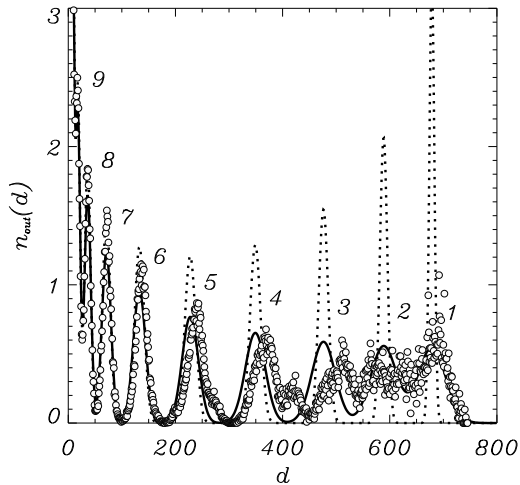}
\caption[]{Comparison of the theoretical out-degree distributions with
numerical data (circles). The dotted line shows the 
theoretical result for a network in the large $L$-limit, where the network
is self-averaging and thus all possible realizations 
of a string of a given length $l$ can be found. The number to the right 
of each peak refers to the node length $l$ that contributes 
predominantly to that peak. The solid line is obtained after correcting
for finite size effects (see text for details). 
In both cases, the locations of the peaks are 
accurately predicted. Note that the results for the large-$L$ limit differ strongly
in their predictions for the width and height of each peak
for small $l$, (large $d$). It is evident that the numerical data
exhibits finite size effects for short nodes, 
$l \lesssim 6$.}
\label{Out1adist}
\end{figure}

Finding the distribution of a sum over a {\em finite} random number $n$ of 
independently distributed random variables $Y$ 
can be readily worked out using moment generating functions (see for
example Feller \cite{Feller}). In the case when both $\tilde{n}_k$ and $Y_{lj}$
are binomial it turns out that the resulting distribution is
binomial again, and we find
\begin{equation}
P(\tilde{X}_{l k} = d | l) = \left( \begin{array}{ll} N \\ d \end{array} \right)
\left[ pq^k p(l,k) \right]^d 
\left[ 1 - pq^k p(l,k) \right]^{N - d },
\label{eqn:pxijfs}
\end{equation}
with mean and variance
\begin{eqnarray}
\left < \tilde{X}_{l k} \right > &=& Npq^k p(l,k) \label{eqn:xiavefs} \\
\tilde{\sigma}^{2}_{l k} &=& Npq^k p(l,k)\left[ 1 - pq^{k}p(l,k) \right].  
\label{eqn:finsig2}
\end{eqnarray} 
Thus Eq.~(\ref{eqn:pxijfs}) is the finite size result replacing Eq.~(\ref{eqn:pxij}), 
which is valid in the large-$L$ limit.
As remarked before, the means of the two
distributions in Eqs.~(\ref{eqn:xiave}) and (\ref{eqn:xiavefs}) are equal,  i.e., 
$\left < \tilde{X}_{lk} \right > = \left < X_{lk} \right >$. However, the variances 
are different and $\sigma^2_{lk} < \tilde{\sigma}^{2}_{lk}$.  
Note that the second term in Eq.~(\ref{eqn:finsig2}) 
is of the order of $(1-p) \approx 1$ for small $p$. Thus we find
to order $p$
\begin{equation}
\tilde{\sigma}^{2}_{l k} = \left < \tilde{X}_{l k} \right >,
\label{eqn:siglkfs}
\end{equation}
and consequently to this order the mean and variance of $\tilde{X}_l$ become
\begin{equation}
\tilde{\sigma}^{2}_{l} = \left < \tilde{X}_l \right > = \left < X_l \right > = d_l,
\label{eqn:sigfinlrel}
\end{equation}
where $d_l$ is the same mean out-degree that was previously obtained in the 
large $L$-limit, Eq.~(\ref{eqn:di}). 
The out-degree distribution corrected for finite-size effects thus becomes, 
c.f., Eq.~(\ref{eqn:poutgauss}),
\begin{equation}
\tilde{P}_{\rm out}(d) = \sum_{l=1}^{L} pq^l 
\frac{1}{\sqrt{2 \pi d_l} } 
\exp \left[ -\frac{ ( d - d_l  )^2}{2 d_l} \right].
\label{eqn:poutfs}
\end{equation}
Comparing this expression with the distribution obtained in the large-$L$ limit, 
Eq.~(\ref{eqn:poutgauss}), we find that finite size corrections are only present for small 
$l$, 
since we have already shown that the relation  $d_l = \sigma^2_l$ is also valid (viz. 
Eq.~(\ref{also})) in the large $l$ region for the  large-$L$ case.
Figure (\ref{Out1adist}) shows a comparison of the numerically obtained 
out-degree distribution (circles) with the theoretical expressions with
and without finite size corrections. The solid line is the 
analytical result for the out-degree distribution,  
Eq.~(\ref{eqn:poutfs}), that takes into account finite size corrections, while
the dotted line corresponds to the case where the network is assumed
to be self-averaging, i.e., Eq.~(\ref{eqn:poutgauss}) is satisfied, and thus sample to 
sample fluctuations can be neglected. 
Note the large difference from the observed behavior for $l < 6$, 
($d > 200$) in the height and 
broadness of the distributions, when finite size effects are not taken 
into account. The agreement of the finite size corrected distribution 
with the numerical data, on the other hand, is rather good, and we conclude 
that finite size effects present in the numerical data for short nodes are satisfactorily 
accounted for.
\vspace*{1cm}

The location of the peaks, $d_l$, coincide very well with the 
numerical data and we find indeed that each peak corresponds to the out-degree
of nodes of a given length $l$. The locations of the peaks decreas exponentially  with
increasing $l$. The labels next to each peak show the string lengths $l$
contributing predominantly to that peak. 

Our reasoning above already
shows that the oscillatory part of the out-degree distribution
is highly succeptible to finite-size effects. 
It turns out that these oscillations are less pronounced or completely absent
when {\em single} finite-size realizations of the network are considered. In other words, 
these oscillations become apparent only when averaging over many
finite-size realizations, as we have done in our analysis. 
   
We turn next to a discussion of the scaling behavior.

\subsubsection{Scaling Behavior}
Our analysis shows that the out-degree distribution is a superposition of
Gaussian peaks with mean $d_l$ and a variance that depends on the strength
of finite size effects, as discussed in the previous section. For
large values of $d$, (small $l$) these peaks are well separated and one
can readily obtain the envelope for the peaks. From Eq.~(\ref{eqn:poutfs})
we see that the height $E_l$ of a 
peak centered at $d_l$ is  
\begin{equation}
E_l = \frac{Npq^l}{\sqrt{2 \pi d_l}}.
\end{equation}
Using Eq.~(\ref{eqn:di}), we obtain the scaling behavior 
\begin{equation}
E(d) \approx d^{-\gamma_2}
\label{eqn:envscale}
\end{equation}
with
\begin{equation}
\gamma_2 = \frac{1}{2}\frac{\ln z - \ln q}
{\ln z + \ln q }.
\end{equation}
For the bit string model with exact matches, i.e., for $r=2$ and in the 
$\beta \to \infty$ limit, we find
\begin{equation}
\gamma_2 = \frac{1}{2}\frac{\ln 2 + \ln q}
{\ln 2 - \ln q}.
\label{eqn:bitscale}
\end{equation}
For the numerical data shown, $q=0.95$, yielding
$\gamma_2 = 0.43$.

For smaller values of $d$ (large $l$), the analysis 
presented above ceases to be valid, since the peaks start
to overlap. In this regime, the contributions to the
out-degree distributions come predominantly from 
matches between long strings which are rare. 
As was remarked previously, in this regime the
distribution of $\tilde{X}_l$ will be Poisson, so 
that we have
\begin{equation}
p(d|l) = \frac{d_{l}^{d}}{d!} e^{-d_l},
\label{eqn:p(d|l}
\end{equation}
with $d_l$ as given before in (\ref{eqn:di}). 
The out-degree distribution for small $d$ is thus given by
\begin{equation}
p(d) = \sum_{l = l^*}^{\infty} pq^l \frac{d_{l}^{d}}{d!} e^{-d_l}
\label{eqn:koutsum}
\end{equation} 
Since for small $l$ the $d_l$ values are quite large, the contributions 
from the small $l$ terms will be suppressed heavily by the exponential 
factor, and therefore moving the cutoff $l^*$ in the above sum down
to $1$ will not change the result of the summation significantly.
Noting that for large $l$
\begin{equation}
d_l = \frac{N}{p} \left (qz \right )^l,
\label{eqn:dilarge}
\end{equation}
we see that $d_{l}$ and $\Delta d_l = d_{l+1} - d_l$ approach
zero in a geometric fashion. Thus the summation over $l$ in 
Eq.~(\ref{eqn:koutsum}), can be converted to an integration
over $x=d_l$ with  $\Delta x = d_{l}-d_{l+1}$ and we obtain
\begin{equation}
p(d) = \frac{c}{d!}\int_{0}^{x^*} x^{d - \gamma_2 - \frac{1}{2}}
e^{-x} dx,
\end{equation}
where $x^* = d_{l^*}$ and $c$ is an overall numerical constant,
\begin{equation}
c=\frac{p}{\ln qz} \left( \frac{N}{p} \right )^{- \frac{1}{2} - \gamma_2}.
\end{equation}
The dominant contribution to the integrand
comes from $x \approx d < x^*$ and we therefore extend the upper limit
to infinity obtaining
\begin{equation}
p(d) = c \frac{\Gamma(d+\frac{1}{2}-\gamma_2)}{\Gamma(d+1)},
\end{equation}
where $\Gamma(x)$ is the gamma function. 
The leading order behavior of $\ln 
\Gamma (x)$ is given asymptotically, for large $x$, by
\begin{equation}
\ln \Gamma(x) =  \left ( x - \frac{1}{2} \right ) \ln x -x + \frac{1}{2} \ln 2\pi + 
O \left (\frac{1}{x} \right ).
\end{equation}
Using the above expansion, we obtain after a little bit of algebra
\begin{equation}
\ln p(d) = {\rm const.} - \left ( \gamma_2 + \frac{1}{2} \right ) \ln d + 
O \left ( \frac{1}{d} \right ).
\end{equation}
It can be readily checked that this approximation for $\ln p(d)$ is 
good even for small values of $d$ and thus p(d) exhibits scaling behavior,
$p(d) \approx d^{-\gamma_1}$, with scaling exponent
\begin{equation}
\gamma_1 = \frac{1}{2} + \gamma_2
\end{equation}

For the numerical data with $z = 1/2$ and $q=0.95$ we find
$\gamma_1 = 0.93$. 

As we have pointed out above, the cross-over between the two
scaling regimes occurs when the depression (minimum) between
consecutive peaks disappears. This occurs roughly when 
\begin{equation}
d_{l+1} + \frac{1}{\sqrt{2d_{l+1}}} > d_l - \frac{1}{\sqrt{2d_l}}
\end{equation}
yielding, via Eq.~(\ref{eqn:di}), 
\begin{equation}
d_l > \frac{1}{2} \frac{1}{1 - \sqrt{1-qz}}.
\end{equation}
For the values of the parameters employed in the numerical simulations, 
this gives $d_l > 6.59$, $\ln d_l > 1.9$, which 
is consistent with the data shown in figure (\ref{outdistscaling}).

We can also infer the large $r$ behavior of  
$\gamma_1$ and $\gamma_2$ for perfect matches.
This corresponds to the case $z=1/r$, eq. (\ref{eqn:zdef}).
We find 
\begin{equation}
\gamma_2 = \frac{1}{2}\frac{\ln r + \ln q}
{\ln r - \ln q},
\label{eqn:rscale}
\end{equation}
and hence
\begin{equation}
\lim_{r \rightarrow \infty} \gamma_2 = \frac{1}{2}
\end{equation}
and correspondingly $ 1/2 + \gamma_2 = \gamma_1 \rightarrow 1$ in this limit. Thus, as the
number of letters in the alphabet is increased, the scaling exponents
$\gamma_1$ and $\gamma_2$, approach the values $1$ and $1/2$, respectively. 
Comparing with the
values for $r=2$, we see that the dependence of $\gamma_1$ and $\gamma_2$ on $r$, 
the number of letters in the alphabet, is rather weak.

\subsection{The In-Degree Distribution}

Consider a randomly selected string $G_i$ of length $l$. Then the random 
variable $X_{k l}$ that was introduced before, counts the number of
edges originating from a string of length $k \le l$ and terminating
in $G_i$. Thus the in-degree of $G_i$ is given by 
\begin{equation}
X_{{\rm in},l} = \sum_{k \le l} X_{k l}.
\label{eqn:xiinsum}
\end{equation}
The statistics of $X_{k l}$ and hence of $X_{in,l}$ has been already 
obtained before and we find in the large-$L$ limit,
\begin{eqnarray}
d_{{\rm in},l} &=&  \sum_{k \le l} n_k p(k,l) 
\label{eqn:xinave}\\    
\sigma_{{\rm in},l}^{2} &=& \sum_{k \le l} n_k p(k,l)\left ( 1 - p(k,l) \right ).
\label{eqn:xinsdev}
\end{eqnarray}
Noting also that the probability of selecting a string of 
length $l$ is $pq^l$, the total in-degree distribution
in the large-$L$ limit is given by
\begin{equation}
P_{\rm in}(d) = \sum_{l=1}^{L} pq^l 
\frac{1}{\sqrt{2 \pi} \sigma_l } 
\exp \left[ -\frac{ ( d - d_{{\rm in},l}  )^2}{2 \sigma_{{\rm in},l}^{2}} \right].
\label{eqn:pingauss}
\end{equation}

When taking into account finite size effects, the in-degree distribution becomes
({\it cf.} Section IV.A.1)
\begin{equation}
\tilde{P}_{\rm in}(d) = \sum_{l=1}^{L} pq^l 
\frac{1}{\sqrt{2 \pi} \tilde{\sigma}_{{\rm in},l} } 
\exp \left[ -\frac{ ( d - d_{{\rm in},l} )^2}{2 \tilde{\sigma}_{{\rm in},l}^{2}} \right],
\label{eqn:pinfs}
\end{equation}
where
\begin{equation}
\tilde{\sigma}_{{\rm in},l}^{2} = \sum_{k \le l} 
Npq^k p(k,l)\left ( 1 - pq^k p(k,l) \right ).
\end{equation}

\vspace*{1cm}
\begin{figure}[h!]
\begin{center}
\end{center}
\includegraphics[width=8cm]{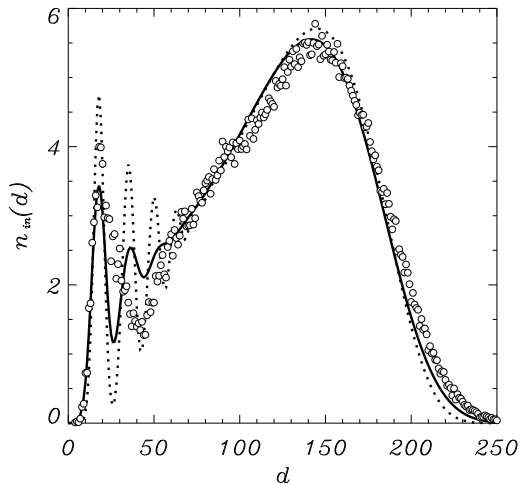}
\caption[]{Comparison of the theoretical in-degree distributions with
numerical data (circles). The dotted line shows the 
theoretical result for a network in the large $L$-limit, where the network
is self-averaging and thus almost all possible realisations 
of a string of a given length $l$ can be found. The solid line is obtained 
after correcting for finite size effects (see text for details). 
}
\label{InDist}
\end{figure}

Unfortunately, we have not been able to obtain closed-form expressions for
$d_{{\rm in},l}$ and $\tilde{\sigma}_{{\rm in},l}$, in a manner analogous to 
the expressions for the out-degree, Eqs.~(\ref{eqn:di}) and 
(\ref{eqn:sigfinlrel}). In the case of the in-degree distributions, 
Eq.~(\ref{eqn:xiinsum}) requires a sum over the first argument of 
the matching probability, $p(\cdot,\cdot;z)$, 
Eq.~(\ref{eqn:pdef2}), rather than the second
argument, as was the case for the out-degree distribution. 
Due to the complicated dependence of the matching probability on 
its first argument this sum is, as far as we can tell, intractable. The
necessary summations were therefore carried out numerically. 
 
Figure (\ref{InDist}) shows a comparison of the two theoretical predictions, 
Eqs.~(\ref{eqn:pingauss}) and (\ref{eqn:pinfs}) with the numerical data of
Balcan and Erzan \cite{Balcan-Erzan}.

The in-degree distribution Eq.~(\ref{eqn:pingauss}), and its finite-size corrected form, 
Eq.~(\ref{eqn:pinfs}), capture the qualitative features seen in the simulations.
Although there are deviations for small and large values of $d$, we will not 
pursue this any further in the present paper. 

Note however the stark difference between the shape of the in- and out-degree distributions, 
Figs.~(\ref{Out1adist}) and (\ref{InDist}). Apart from the distinct qualitative
features, such as  oscillatory behavior for small $d$ 
(rather than large $d$ as in the out-degree distribution), 
the in-degree distribution is much narrower than the out-degree distribution.

\section{Discussion}

We have obtained analytical expressions for the in- and out-degree distribution
of a contents-based network model which was introduced 
and studied numerically by Balcan and Erzan in~\cite{Balcan-Erzan}. 
We have shown that the behavior of the 
out-degree distribution can be divided into two regimes:  
a short and putative scaling regime for small out-degrees that crosses over into
an oscillatory regime for large out-degrees. An analytical expression for the 
cross-over point has been obtained as well. We have found that the behavior of the
out-degree distribution for large out degrees depends on the size of the network 
realizations from which the distribution was sampled. We have discussed these 
finite-size effects and have shown analytically how they effect the 
behavior of the out-degree-distribution. 

Our results were obtained for a generalized class of contents-based network models
in which a small number of imperfect matches (finite, but low, temperature) were 
allowed and strings were constructed from an alphabet of $r$ letters. 
It turns out, however, that such generalization do not alter the main numerical 
findings of the network model of Balcan and Erzan which involved   
a two-letter ($r=2$) alphabet and perfect matches.  The scaling behavior which we have 
found, and even the numerical values of the leading scaling exponents 
$\gamma_2$ and $\gamma_1=\gamma_2+ 0.5 $ 
are robust under these generalisations.  It should be noted that, in 
\begin{equation}
\gamma_2 = \frac{1}{2}\frac{\ln z - \ln q}
{\ln z + \ln q }\;\;,
\end{equation}
we have $z \to 1/r$ for $\beta \to \infty$, while $z \to 1$ in the ``high temperature'' 
limit $\beta \to 0$, thus $r^{-1} \le z \le 1$.  In the ``low temperature,'' or perfect 
matching, limit  $\beta \to \infty$, 
\begin{equation}
\gamma_2 \to (1/2) (1-p/ \ln r), 
\end{equation}
where $p$ is a small number by assumption~\cite{Balcan-Erzan}.
Even when allowing for a small number of mismatches,
$\gamma_2$ depends very weakly on $r$, and $p$. On the other 
hand, for either $r\to 1$, the trivial limit where no information is coded, or the high 
temperature limit, where no matching conditions are satisfied, the scaling relation is 
altered qualitatively, with $\gamma_2 \to  -1/2$.

We should remark on the robustness of the incipient power 
law behavior found in the limit of  small degrees, for the out-degree 
distribution. Two different sources of randomness determine 
together the degree distributions of our model through the variables 
$d_l$ and $n_l$. While $d_l$ is determined by the adjacency rule based on 
sequence matching, and therefore 
depends on the length of the sequence to be matched, the distribution of 
nodes of length $l$ could have been chosen in many different ways.  
The exponential dependence turns out to be algebraically tractable, 
but it may be conjectured that any distribution which 
has a tail that is decaying exponentially with $l$ would give rise, 
all else remaining equal, to essentially the same scaling behavior for the 
out-degree distribution in the large $l$ (small $d$) regime, and therefore 
that $\gamma_1 \simeq 1$ has a high degree of universality.

We would like to end this paper by pointing out the possible 
relevance of this network model to understanding molecular networks~\cite{Mirni}, 
in particular transcriptional genomic networks. 

Transcriptional genomic networks are obtained by identifying the nodes with genes, 
and the directed edges connecting two nodes with so called transcription 
factors (TF). A TF is the protein coded by the gene at the node of origin,  
and binds (i.e., becomes chemically attached to) a short DNA sequence within 
the promoter region typically upstream of the target gene, whose activity it 
controls by either promoting, or suppressing it.~\cite{genenetworks,biobook2}

An assay of the recently available results coming from high-throughput 
experiments on the degree distribution of transcriptional genomic networks 
reveals that the out-degree distribution shows putative scaling over 
a very short range of about one decade at most, with a lot of scatter, 
and a marked departure from linearity on double logarithmic plots, 
for larger degrees. Nevertheless, with the assumption that $n(d)\sim d^{-\gamma}$ 
over the whole range, the exponents $\gamma$ which have been reported 
are all smaller than two, and closer to unity: $\gamma = 1.4$ (yeast) \cite{Lee},
$\gamma = 1$ (yeast) \cite{Guelzim}, $\gamma = 1.1$-$1.8$ (several genomes) 
\cite{Barkai}, $\gamma = 1.5$ ({\it E. coli}) \cite{Dobrin},
$\gamma = 1.3$ (yeast) \cite{Tong}.

Comparing these findings for the degree distribution of the 
transcriptional regulatory network with the results of our model is very suggestive. 
The marked but short range over which the data can indeed be fitted by 
a straight line in a log-log plot of the degree distribution has a power 
close to unity, as found in the experiments on 
transcriptional regulatory networks cited above. The crossover 
to a different regime  towards the tail end of the distribution, is a 
feature that also shows similarity with the  experimental results.
Clearly the oscillations of the out-degree distribution, 
Figs.~(\ref{outdistscaling}) and (\ref{Out1adist}) 
are not seen in the degree distributions of the transcription regulatory 
networks extracted from any particular genome. In the language 
of our paper, real cellular networks are more like single finite-size realizations, 
rather than expected distributions calculated over ensembles of many different 
realizations of a random sequence.  In our model, for any particular finite-size 
realization, only a relatively small number of data points would fall into this 
portion of the distribution and this would not be sufficient to resolve well the 
oscillations that make up the sample-averaged distribution. The small 
degree behavior of the degree distribution, however, is robust 
with respect to sample-to-sample fluctuations, as we have shown. 

We think that the similarity  with reported degree statistics 
of transcriptional genomic networks is not fortuitous. 
Sequence matching provides a highly plausible mechanism for the formation of 
the transcriptional regulatory network. Such networks rely on the recognition 
of regulatory sequences by transcription factors.  
These points will be discussed in detail within a more comprehensive comparison 
of features of content-based network models with real biological data in a 
forthcoming article~\cite{Alkan}.

\vskip 1cm
{\bf Acknowledgements}

One of us (AE) gratefully acknowledges partial support from the Turkish Academy of Sciences.
MM gratefully acknowledges partial support from the Nahide and Mustafa Saydan Foundation.
AK acknowledges support of FIRB01.
\vskip 2cm
\appendix

{\bf APPENDIX}
\nopagebreak
\smallskip

Here we outline the calculations leading to Eqs. (\ref{eqn:fij2})
and (\ref{eqn:factorn}). In Section III, we defined the function 
$W^{(2)}(a,b;{\bf x})$ as, Eq.~(\ref{eqn:meierterms}),
\begin{equation}
W^{(2)}(a,b;{ \bf x}) = \frac{1}{r^k} 
\sum_{{\bf y}}^{}f_a({\bf x},{\bf y};\beta)f_b({\bf x},{\bf y};\beta)
\label{eqn:meiertermsa}
\end{equation}

As we pointed out in the text, when performing the sum over ${\bf y}$, two
cases must be distinguished: (i) $|b - a| \ge l$ and (ii) $|b-a| < l$.
In case (i), the set of indices of ${\bf y}_{a,l}$ and ${\bf y}_{b,l}$ are 
distinct and the evaluation of the partition sum proceeds in a manner 
analogous to Eq.~(\ref{eqn:fi}) yielding
\begin{equation}
W^{(2)}(a,b;{\bf x}) = 
\left (\frac{1}{r^l} \right )^2 \left [ 1 + (r-1)e^{-\beta} \right ]^{2l}, \, 
|b-a| \ge l.  
\label{eqn:fij1a}   
\end{equation}

\begin{figure}[h!]
\begin{center}
\end{center}
\includegraphics[width=8cm]{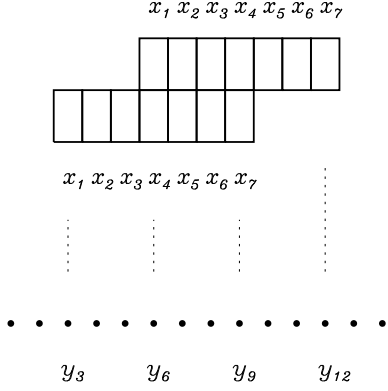}
\caption[]{Schematic representation of the $y$ and $x$ averages for the 
function $W^{(2)}(a,b;\beta)$ as defined in the text. The case shown 
in the figure corresponds to 
$l = 7$ with $a=2$ and $b=5$. The number of overlapping indices in the figure
is $l-m(=4)$, with $m= b-a(=3)$. Because of the overlapping, when averaging over 
${\bf x}$,  these indices fall into $m(=3)$ disjoint sets: $\{ x_1,x_4,x_7 \}$,
$\{ x_2,x_5 \}$ and $\{ x_3,x_6 \}$ }.
\label{boxfig}
\end{figure}

In case (ii) there is an overlap between the indices of ${\bf y}_{a,l}$ 
and ${\bf y}_{b,l}$. Defining $|b-a| = m$, we find that there are 
$l-m$ overlapping indices, and thus there are $k - (l + m)$ distinct 
variables $y_c$ that are neither in ${\bf y}_{a,l}$ nor in ${ \bf y}_{b,l}$, 
so that a sum over the values of these indices will give $r^{k-(l + m)}$. 
Next, it is convenient to partition the remaining indices, 
$\{y_{a+1}, \ldots, y_{b+l}\}$, into the three disjoint sets, 
$S_1 = \{y_{a+1}, \ldots , y_{a+m}\}$, 
$S_2 = \{y_{a+m+1} = y_{b+1}, \ldots, y_{a+l} = y_{b+l-m+1}\}$ and
$S_3 = \{y_{b+a-m+2}, \ldots , y_{b+l}\}$. 
Figure (\ref{boxfig}) shows an example for $l=7$, with $a=2$ and $b=5$ along
with the sets, $S_1 = \{y_3,y_4,y_5\}$, $S_2 = \{y_6,y_7,y_8,y_9\}$ and
$S_3 = \{y_{10},y_{11},y_{12}\}$. 
With the definitions above, we find for $|b-a| < l$,
\begin{eqnarray}
&W&^{(2)}(a,b;{\bf x}) =   
\frac{1}{r^{l+m}} \sum_{S_1} e^{-\beta \sum_{t=1}^{k} u(x_t,y_{a+t})} \nonumber \\ 
       &\times& \sum_{S_3} e^{-\beta \sum_{t=1}^{m} u(x_{b+l-m+t}),
y_{b+l-m+t})} \\ &\times& \sum_{S_2} 
e^{-\beta \sum_{t=1}^{l-m} \left [ u(x_t,y_{b+t}) + u(x_{m+t},y_{b+t}) \right ]} \nonumber
\end{eqnarray}
and carrying out the sums over the $y$ variables, we obtain ( $|b-a| < l$),
\begin{eqnarray}
W^{(2)}(a,b;{\bf x}) =   
\frac{1}{r^{l+m}} \left [ 1 + (r-1)e^{-\beta} \right ]^{2m} \nonumber \\
\times \prod_{t=1}^{l-m}
   \left [ 1 + (r-1) e^{-2\beta} - u(x_t,x_{t+m})\left ( 1 - e^{-\beta } \right )^2 
   \right ].
\label{eqn:aafij2}
\end{eqnarray}

\vspace*{1cm}

Next, it is useful to introduce the $r \times r$ matrix, $M(x,y)$ as
\begin{equation}
M(x,y) =  1 + (r-1) e^{-2\beta} - u(x,y)\left ( 1 - e^{-\beta } \right )^2, 
\end{equation}
with $x,y \in \{ 0,1,2, \ldots, r-1 \}$.
From the properties of $u$, Eq.~(\ref{eqn:udef}), we find that
\begin{equation}
M(x,y) = \left \{ \begin{array}{ll}   1 + (r-1) e^{-2\beta}, & x = y \\
                                      (r-2) e^{-2\beta} +2 e^{-\beta},   & x \ne y 
\end{array} \right.
\label{eqn:mdef}
\end{equation}
and Eq.~(\ref{eqn:aafij2}) can therefore be written as
\begin{eqnarray}
W^{(2)}(a,b;{\bf x}) &=&   
\frac{1}{r^{l+m}} \left [ 1 + (r-1)e^{-\beta} \right ]^{2m} \nonumber \\
&\times& \prod_{t=1}^{l-m} M(x_t,x_{t+m}). 
\label{eqn:aafij3}
\end{eqnarray}
Proceeding to perform the average over ${\bf x}$,
\begin{equation}
W^{(2)}(a,b) = \frac{1}{r^l} \sum_{{\bf x}} W^{(2)}(a,b;{\bf x}),
\label{eqn:aafijave}
\end{equation}
observe that in eq. (\ref{eqn:aafij3}) the variables  ${\bf x}$ can
be partitioned into $k$ disjoints sets $X$ with the additional 
property that if $x_t \in X$, by implication $x_{t+m} \in X$. 
The situation is shown schematically in Fig. (\ref{boxfig}) 
for $m=3$, where we have the $3$ disjoint sets, $\{ x_1,x_4,x_7 \}$, 
$\{ x_2,x_5 \}$ and $\{ x_3,x_6 \}$.
Denoting these sets as $X_1, X_2, \ldots X_m$, and their 
respective number of
elements as $n_1, n_2, \ldots , n_m$ ($n_1 + n_2 + \ldots n_m = l$),
we see that the product in Eq.(\ref{eqn:aafij3}) can be factorized as
\begin{equation} 
\prod_{t=1}^{l-m} M(x_t,x_{t+m}) = \prod_{x_t \in X_1 } M(x_t,x_{t+m})
 \cdots \prod_{x_t \in X_m } M(x_t,x_{t+m}) 
\end{equation}
Performing the summation over each of the factors we have for the 
first factor 
\begin{equation}
\sum_{X_1} \prod_{x_t \in X_1 } M(x_t,x_{t+m}).
\end{equation}
It can be easily shown that the sum over the variables $x_t \in X_1$ reduces to an
$n_1-1$ fold matrix product. Denoting the matrix elements of the matrix
$M^{n}$ by ${(M^n)}_{xy}$, we therefore find
\begin{equation}
\sum_{X_1} \prod_{x_t \in X_1 } M(x_t,x_{t+m}) = \sum_{x,y} {(M^{n_1-1})}_{xy} 
\end{equation}
and hence
\begin{equation}
\frac{1}{r^l} \sum_{\bf x} \prod_{t=1}^{l-m} M(x_t,x_{t+m}) = 
\frac{1}{r^l} \prod_{s=1}^{m}\sum_{x,y} {(M^{n_s-1})}_{xy} 
\label{eqn:xave1}
\end{equation}

Owing to the structure of the matrix $M$, Eq.~(\ref{eqn:mdef}), 
powers of $M$ retain the same structure, as can be readily shown, and we 
therefore have
\begin{equation}
{(M^{n})}_{(xy)} = \left \{ \begin{array}{ll}   A_n, & x = y \\
                                      B_n,   & x \ne y. 
\end{array} \right.
\label{eqn:mdefn}
\end{equation}
The quantities $A_n$ and $B_n$ can be evaluated  recursively, and one
finds after a little algebra,
\begin{equation}
\left ( \begin{array}{l} A_{n+1} \\ B_{n+1} \end{array} \right ) 
= Q_n 
\left ( \begin{array}{l} A_1 \\ B_1 \end{array} \right ),
\end{equation}
where
\begin{equation}
Q_n = 
\frac{1}{r} \left ( \begin{array}{cc} (r-1)\lambda_{+}^{n} + \lambda_{-}^{n} 
& - (r-1)\lambda_{+}^{n} + (r-1) \lambda_{-}^{n} \\
- \lambda_{+}^{n} + \lambda_{-}^{n} & \lambda_{+}^{n} + (r-1) 
\lambda_{-}^{n} \end{array} \right ),
\end{equation}
and
\begin{eqnarray}
\lambda_{+} &=&  \left [ 1 - e^{-\beta} \right ]^2 \\
\lambda_{-} &=&  \left [ 1 + (r-1)e^{-\beta} \right ]^2 .
\end{eqnarray}

We therefore find,
\begin{equation}
\sum_{x,y} {(M^{n})}_{xy} = r A_n + r(r-1)B_n,
\end{equation} 
and thus
\begin{equation}
\sum_{x,y} {(M^{n})}_{xy} = r \left [ 1 + (r-1)e^{-\beta} \right ]^{2n}.
\label{eqn:mntrace}
\end{equation}
Substituting eq. (\ref{eqn:mntrace}) back into eq. (\ref{eqn:xave1}) we have
\begin{equation}
\frac{1}{r^l} \sum_{\bf x} \prod_{t=1}^{l-m} M(x_t,x_{t+m}) = 
\frac{1}{r^l} \prod_{s=1}^{m} r \left [ 1 + (r-1)e^{-\beta} \right ]^{2(n_s-1)} 
\end{equation}
and noting that $n_1 + n_2 + \ldots n_m = l$, we finally obtain
\begin{equation}
\frac{1}{r^l} \sum_{\bf x} \prod_{t=1}^{l-m} M(x_t,x_{t+m}) = 
\frac{1}{r^l} r^m \left [ 1 + (r-1)e^{-\beta} \right ]^{2(l-m)}, 
\end{equation}
which when substituted into Eqs.~(\ref{eqn:aafijave}) and (\ref{eqn:aafij3}) 
yields the final result, Eq.~(\ref{eqn:factor1}),
\begin{equation}
\frac{1}{r^l} \sum_{{\bf x}} W^{(2)}(a,b;{\bf x}) = 
  \frac{1}{r^{2l}} \left [ 1 + (r-1) e^{-\beta} \right ]^{2l}.
\end{equation}

Note that we obtain the same result as for the case $|b-a| > l$, Eq.~(\ref{eqn:fij1a}). 
In particular, we see that once averaged over ${\bf x}$, $W^{(2)}$ is independent of $a$ and
$b$.

\end{document}